\documentclass[conference]{IEEEtran}

\usepackage{amssymb}
\usepackage{amsmath}

\usepackage{amsthm,theorem}
\usepackage{amssymb}
\usepackage{graphicx,tikz}
\usetikzlibrary{arrows,automata}

\usepackage{array,cite}
\usepackage{subfigure,epsfig}

\title{Molecular Communication Between Two Populations of Bacteria}
\author{  Arash Einolghozati, Mohsen Sardari, Faramarz Fekri\\
School of Electrical and Computer Engineering\\
Georgia Institute of Technology, Atlanta, GA 30332\\
\texttt{Email:}\{einolghozati, mohsen.sardari, fekri\}@ece.gatech.edu
\thanks{This material is based upon work supported by the National Science Foundation under Grant No. CNS-111094}
}

\begin{document}
\IEEEoverridecommandlockouts
\maketitle

\begin{abstract}
Molecular communication is an expanding body of research. Recent advances in biology have encouraged using genetically engineered bacteria as the main component in the molecular communication. This has stimulated a new line of research that attempts to study molecular communication among bacteria from an information-theoretic point of view. Due to high randomness in the individual behavior of the bacterium, reliable communication between two bacteria is almost impossible. Therefore, we recently proposed that a population of bacteria in a cluster is considered as a node capable of molecular transmission and reception. This proposition enables us  to form a reliable node out of many unreliable bacteria. The bacteria inside a node sense the environment and respond accordingly. In this paper, we study the communication between two nodes, one acting as the transmitter and the other as the receiver. We consider the case in which the information is encoded in the concentration of molecules by the transmitter. The molecules produced by the bacteria in the transmitter node propagate in the environment via the diffusion process. Then, their concentration sensed by the bacteria in the receiver node would decode the information. The randomness in the communication is caused by both the error in the molecular production at the transmitter and the reception of molecules at the receiver. We study the theoretical limits of the information transfer rate in such a setup versus the number of bacteria per node. Finally, we consider M-ary modulation schemes and study the achievable rates and their error probabilities. 
\end{abstract}

\section{Introduction}
\label{sec:intro}
 
The use of bacteria as means of communication is inspired from naturally occurring communication between bacteria through a process called Quorum Sensing (QS). Molecular communication between bacteria is conducted in such a way that a population of bacteria can reliably infer information about their environment~\cite{Bassler1999}. Bacteria use molecules to exchange information among themselves to be able to perform a task otherwise impossible~\cite{kaplan1985,Bassler1999}. Some examples for this coordinated task are light production and attacking the host by bacteria. In QS, each individual bacterium in a population releases specific types of molecules to the environment. The concentration of molecules in the environment (sensed by the same population of bacteria) is a measure of the local density of bacteria. Bacteria performs their task when the concentration of molecules surpasses a threshold.


New applications and designs are constantly emerging from manipulation of the genetic content of QS bacteria. In~\cite{Tamsir2011}, a simple genetic circuit is used with QS in order to design logical gates, i.e., the output of bacterium coincides with a logical table according to presence or absence of specific molecules in its vicinity. QS is used in~\cite{Danino2010} to design biological clocks, i.e., regulation of the output of a population of bacteria to alternate periodically. There has been also new research in network engineering inspired by this phenomenon. For example, models for forming a network via molecular communication are given in~\cite{INFOCOM2012_Arash,Akyildiz2011}. In these studies the information is encoded in the concentration of molecules. This information model departs from another line of research which relies on encoding the information in the \emph{timing} of emission of molecules~\cite{eckford,rose2011}. All these studies have inspired researchers to investigate the communication among bacteria more carefully and also pay attention to information-theoretic aspects of bacteria communication~\cite{ISIT2011_Arash, ITW2011_Arash}.

The communication between bacteria is slow and the number of bits transferred is small. On the other hand, the reliability of the communication due to large number of bacteria in the environment can be significant. The main motivation is to enable reliable communication in the networks that are bio-compatible as well. These networks have sensory applications and the delay in the communication can be fairly large.

\subsection{Problem Setup}

As shown by the previous studies in biology~\cite{perez2010}, the individual behavior of bacteria has a heterogeneous nature and may involve high levels of randomness. In order to form reliable communication out of unreliable bacteria, we consider the communication between a population of bacteria residing in a node. In~\cite{INFOCOM2012_Arash}, we introduced a Molecular Communication Networking (MCN) paradigm where populations of bacteria (i.e., the primitive agents), clustered together and acted harmonically, form a node in a communication network. Each bacterium in a node is able to produce molecules, sense the concentration of molecules (from a chemical substance) in the environment and respond accordingly. This particular response is programmed into the plasmid which is embedded in the bacteria to act along with the DNA of bacteria. Such a node in MCN is considered to be an independent entity and act as a fairly smart node in the network. In such proposed networks, the communication happens between the nodes instead of the individual bacteria. The sensed information is relayed in the network from one node to another through diffusion of molecules in the medium. This setup enables us to take advantage of the primitive agents (i.e., the engineered bacteria) in a network that is designed to perform a specific task and transfer information. The information traveled in the network is a specif parameter of the environment, (e.g., existence of a chemical substance and/or its amount in the environment).

\begin{figure}
\includegraphics[width= .48\textwidth]{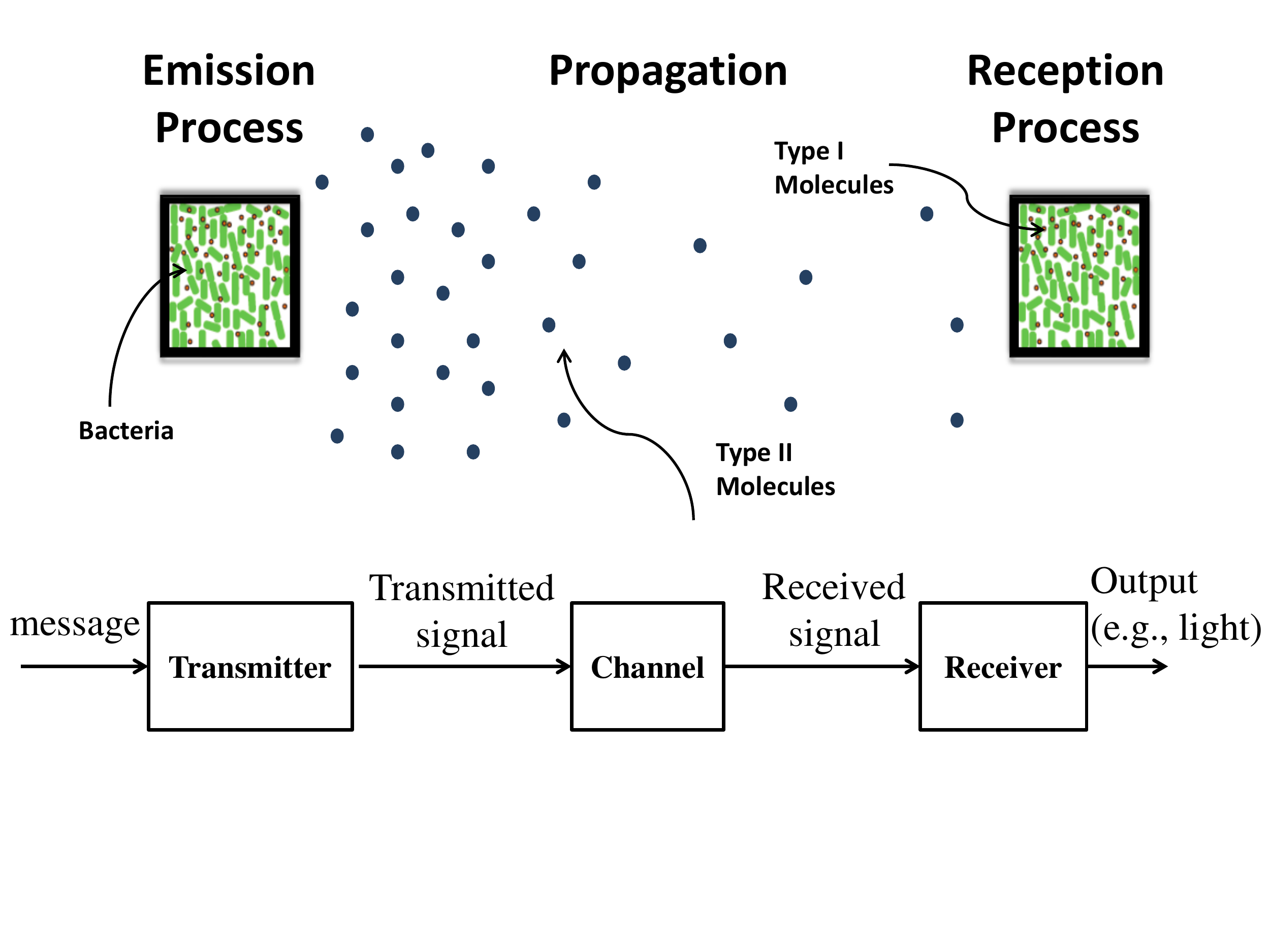}
\vspace{-.3in}
\caption{The molecular communication setup consisting of the transmitter, channel and the receiver}
\vspace{-.1in}
\label{fig:two_node}
\end{figure}

Our goal is to model and analyze the molecular communication between two nodes in a network described above. Toward this, in this paper, we modify our problem to a two-node communication in which one node acts as the transmitter and the other as the receiver (depicted in Fig.~\ref{fig:two_node}. As in~\cite{INFOCOM2012_Arash}, the transmitter node is assumed to be smart enough to stimulate its bacteria to emit molecules into the environment.  The bacteria are stimulated with type I molecules with a proper concentration. These molecules are trapped by the ligand receptors of the transmitter bacteria. Upon this, each bacterium produces type II molecules with concentration rate that depends on the number of its activated receptors. We consider a probabilistic model for the reception of type I molecules which in turn results in a  probabilistic model for the type II output rate. The produced type II molecules by the transmitter travel through the diffusion channel and reach the receiver node. 

The process of reception of type II molecules by the bacteria of the receiver node is similar to that of type I in the transmitter node. The receptors of the bacteria in the receiver node are designed to trap the type II molecules. The difference is that the final output by each bacterium due to reception of type II molecules will be in the form of light or Green Fluorescent Protein (GFP). The receiver node infers the transmitted information from the aggregate output of all the bacteria within the node. We assume the diffusion channel to be noise free; hence, the stochastic nature of the output is due to two factors: the error in the channel input concentration (i.e., the transmission noise) and the error in the reception of molecules at the receiver. Here, we intend to study the maximum rate of information exchange and also analyze M-ary signaling schemes with their rates of error. In~\cite{ISIT2012_Arash}, we described the reception of molecules by the bacteria. In this paper, we extend that model to the transmitter and study the role it plays alongside of the receiver. This further enables us to study the capacity of the molecular communication between two nodes.

The rest of the paper is organized as follows. In Sec.~\ref{sec:transmitter}, the process of production of molecules at the transmitter side is discussed. Sec.~\ref{sec:receiver} studies both the receiver and the achievable rates. Then, Sec.~\ref{sec:modulation} introduces a  practical signaling scheme for the communication setup. Finally, Sec.~\ref{sec:conclusion} concludes the paper.

\section{Transmitter Model}
\label{sec:transmitter}

Let assume the transmitter node $T$ would like to send a concentration $A_0$ to the receiver node $R$. The chamber of the transmitter stimulates the $n$ bacteria (it contains at the node) with type I molecules which in turn produce type II molecules by the bacteria. These molecules would then diffuse through the channel to the receiver. To perform the transmission functionality, each bacterium must be able to receive and decode the type I molecules, emitted by the chamber as stimulus to the node. Each bacterium is assumed to have separate receptors for different types of molecules. We assume $N$ ligand receptors for each type of molecules. Furthermore, the model of the two type receptors is assumed to be the same, i.e. the process of reception follows the same set of equations. To generate the desired type II concentration $A_0$, the agents are stimulated with type I molecules with the appropriate concentration $A_1$ which will be determined later in this paper. 

In order to account for the production and reception of molecules, we use the model introduced in~\cite{muller2008}. This model considers a chain of linear differential equations that account for the output of bacteria in response to presence of molecules in the medium.
In this model, each cell receptor (i.e., the ligand receptor) is  activated with a probability that depends on the concentration of molecules in the medium surrounding the cell. As shown in~\cite{muller2008}, the binding probability $p$ at the steady state is given by
\begin{equation}
\label{eq:steady_state}
p=\frac{A\gamma}{A\gamma+\kappa},
\end{equation}
where $A$ is the concentration observed by the bacterium, $\gamma$ is the input gain and $\kappa$ is the dissociation rate of the trapped molecules from the cell receptors.  
 The process of the production of complex molecules, transcription of genes and the process of the  production of the output are modeled similarly~\cite{muller2008}. The output of bacteria in the steady state is a linear function of the number of the activated receptors.Hence,  the output noise of the bacterium is caused by the probabilistic nature of the ligand reception process~\cite{ITW2011_Arash}.

We assume the noise  in the transmitter output is originated from the discrepancy in the individual behavior of the bacteria in the transmitter node $T$. In other words, even though the average behavior of bacteria can be formulated with a set of deterministic differential equations, the individual behavior of bacteria features randomness. Such randomness can be accounted for by considering the constants in~(\ref{eq:steady_state}) as random variables. Two factors contribute to the uncertainty of the molecular concentration output of a transmitter node. One is the probabilistic nature of the number of activated receptors. We model this by assuming each receptor being active as a Bernoulli random variable that is $1$ with probability $p$ defined in~(\ref{eq:steady_state}). The other factor is the randomness in $p$ itself from one bacterium to another within the node. This is due to the variability of the input gain $\gamma$ in (\ref{eq:steady_state}) within the population of bacteria.
 We model this variation in the input gain $\gamma$ as an iid additive noise $\epsilon_{\gamma}$.  Hence, the entrapment probability $p_1$ upon the reception of the concentration $A_1$ by bacteria would be given by
\begin{equation}
\label{eq:noisy_probability}
p_1=\frac{A_1(\gamma+\epsilon_{\gamma})}{A1(\gamma+\epsilon_{\gamma})+\kappa},
\end{equation}
where  $\epsilon_{\gamma}$ is a zero-mean Normal noise with variances $\sigma_{\gamma}^2$. The variance is assumed to be sufficiently small such that we can ignore the second and higher orders of $(\frac{\epsilon_{\gamma}}{\gamma})$. We assume the same $p_1$ for all the receptors of a bacterium, but it varies according to $\epsilon_{\gamma}$ in~(\ref{eq:noisy_probability}) for different bacteria in a node.

The exact analysis using the expression in~(\ref{eq:noisy_probability}) would be cumbersome. Hence, we only consider the first order terms of $\frac{\epsilon_{\gamma}}{\gamma}$. We define the noiseless input probability as
\begin{equation}
\label{eq:noiseless_probability}
 p_1^*=\frac{A_1 \gamma}{A_1 \gamma+\kappa}.
\end{equation}
By approximating~(\ref{eq:noisy_probability}), we will have
\begin{equation}
\label{eq:noisy_probability2}
p_1=p_1^*+\frac{p_1^*(1-p_1^*)}{\gamma} \epsilon_{\gamma}.
\end{equation}
The total number of activated receptors of $i^{th}$ bacterium, $X_i$, is a Binomial random variable with parameters $(N,p_{1,i})$ where $p_{1,i}$ is the realization of $p_1$ for the $i^{th}$ bacterium. Recall that $N$ is the number of ligand receptors per bacterium for a given molecule type. We denote $X$ as the total number of activated receptors of all bacteria in the node $T$. Hence, $X=\sum_{i=1}^n X_i $. 
 Using the conditional expectation, we have
\begin{equation}
\label{eq:expected_value}
E(X_i)=E(E(X_i|p_{1,i}))=E(Np_{1,i})=Np_1^*, \nonumber
\end{equation}
where the last equality is due to the fact that the noise $\epsilon_{\gamma}$ has zero mean. Hence, we have E(X)=$n N p_1^*$.  %
By using the conditional variance, we have
\vspace{-.05in}
\begin{align}
\label{eq:variance}\nonumber
Var(X_i) &=E(Var(X_i|p_{1,i}))+Var(E(X_i|p_{1,i}))\\
	\nonumber&=E(N p_{1,i}(1-p_{1,i}))+Var(Np_{1,i})\\
&=N p_1^*(1 -p_1^*)+(N^2-N) {p_1^*}^2 (1-p_1^*)^2 \frac{\sigma_{\gamma}^2}{\gamma^2}.
\end{align}
The first term in~(\ref{eq:variance}) is due to the general uncertainty in a Binomial output (i.e., the probabilistic nature of the ligand reception) and the second term is due to the noise in the parameter $p_1$. By independent assumption between the outputs of different bacteria, the variance of the total output by the transmitter node is obtained as
\begin{equation}
\label{eq:variance_output}
 Var(X)=n N p_1^*(1 -p_1^*)+ n(N^2-N) {p_1^*}^2 (1-p_1^*)^2 \frac{\sigma_{\gamma}^2}{\gamma^2}. 
\end{equation}
Since the number of receptors $N$ per bacterium is usually large enough, the second term is dominating. Hence, we can approximate the variance by $n N^2  {p_1^*}^2 (1-p_1^*)^2 \frac{\sigma_{\gamma}^2}{\gamma^2}$. 

As discussed above, the production output of bacteria depends linearly on the number of activated receptors $X$. Hence the total type II molecule output of the node $T$ is equal to $\alpha X$ where $\alpha$ is a constant. The produced molecules are transferred through the diffusion channel. Hence, the steady-state concentration $A_2$ at $R$ will be
\begin{equation}
\label{eq:concentration_R}
A_2=G(r)\alpha X,
\end{equation}
where as shown in~\cite{random_walk}, $G(r)=\frac{1}{4\pi Dr}$ for the ideal channel model. Here $r$ is the distance between the transmitter and the receiver nodes and $D$ is the diffusion coefficient. Moreover, from~(\ref{eq:concentration_R}), we obtain
\begin{equation}
\label{eq:expected_value2}
 E(A_2)=\alpha G(r) n N \frac{A_1 \gamma }{A_1 \gamma+\kappa}.
\end{equation}
The required stimulating concentration $A_1$, can be obtained by putting the right term in~(\ref{eq:expected_value2}) equal to $A_0$; the desired concentration to be transferred from node $T$ to $R$. Hence, we have
\begin{equation}
A_1=\frac{\kappa A_0}{\gamma (\alpha G(r) n N-A_0)}.\nonumber
\end{equation}

In order to make the analysis of the receiver tractable, we approximate the concentration  in~(\ref{eq:concentration_R}) with a Normal random variable. Since the number of receptors $N$ is large, we can use the Central Limit Theorem to approximate $X_i$ by ${\bf \mathcal{N}} (Np_1^*,Var(X_i))$ where $Var(X_i)$ is given in~(\ref{eq:variance}). Hence, the transmitter output $X$ (without including $\alpha$) would  be the sum of $n$ Normal variables given by
\begin{equation}
\label{eq:normal_output} 
X=nNp_1^*+\epsilon_X,
\end{equation}
where $\epsilon_X$ has a $\mathcal{N}(0,Var(X))$ distribution. Hence, the concentration $A_2$ at the receiver would be
\begin{equation}
\label{eq:noisy_concentration}
A_2=A_0+ \epsilon_t.
\end{equation}
where $\epsilon_t$ is a zero-mean Normal random variable with variance $\sigma_t^2=G^2(r) \alpha^2 Var(X)$. The first term in~(\ref{eq:noisy_concentration}) can be viewed as the signal to be decoded by the receiver node and the second term is an additive Gaussian noise which has a signal-dependent variance. In other words, the transmitter induces a concentration $A_2$ of type II molecules at the receiver node which has the desired concentration $A_0$ plus the noise We refer to this noise as the transmitter noise perceived at the receiver in the molecular communication.

\section{ The Receiver and Capacity Analysis} 
\label{sec:receiver}
The concentration $A_2 $ derived in~(\ref{eq:noisy_concentration}) is sensed by the bacteria in the receiver node $R$. The sensing process of type II molecules is similar to that of the type I molecules we analyzed in the transmitter. Hence it follows the equations in the previous section. The difference is that the input concentration is noisy itself which introduces additional uncertainty to the output of the node $R$; which is in the form of light or GFP. This output is the indication of the decoded information sent to R.

Here, we incorporate the effect of both noises introduced in the last section.  Again, the noise $\epsilon_{\gamma}$ accounts for the dependency of gain $\gamma$ on the bacterium at the receiver node and $\epsilon_t$ accounts for the concentration noise introduced in~(\ref{eq:noisy_concentration}).  Hence, the entrapment probability at the receiver can be written as
\begin{equation}
\label{eq:noisy_probability1}
p_2=\frac{(A_0+\epsilon_t)(\gamma+\epsilon_{\gamma})}{(A_0+\epsilon_t)(\gamma+\epsilon_{\gamma})+\kappa}.
\end{equation}
Note that the input concentration noise $\epsilon_t$ affects all the bacteria in the same manner.
By approximating~(\ref{eq:noisy_probability1}) and again keeping only the first order terms of the noises, we obtain
\begin{equation}
\label{eq:noisy_probability2}
p_2=p_0+ \frac{p_0(1-p_0)}{\gamma} \epsilon_{\gamma} +p_0(1-p_0) \epsilon_t,
\end{equation}
where we define $p_0\triangleq \frac{A_0\gamma}{A_0\gamma+\kappa}$. The first term in the right hand side of~(\ref{eq:noisy_probability2}) is due to the actual channel input, the second term is the noise due to the reception process at the receiver (i.e., the gain $\gamma$ varies among the different bacteria in the node $R$) and the third term is contributed by the (transmitter) noise in the receiver input concentration $A_2$. 
We denote by $Y_i$  as the output of $i^{th}$ bacterium in the node $R$. Then, $Y=\sum_{i=1}^n Y_i$ would give the aggregate output of all the $n$ bacteria in the node. For the rest of the discussion, we assume that the output of the node $R$ is in the form of light~\cite{kaplan1985}. Note that $Y$ is the sum of binomial random variables with parameters $(N,P_{2,i})$. Here, $p_{2,i}$ is the realization of $p_2$ for the $i^{th}$ bacterium.

The expected value of the output light can be obtained similar to the transmitter, that is $E(Y)=Nnp_0$. Computing the variance of the output will be more involved. Since $\epsilon_t$ is the same for all the bacteria of a node, $Y_i$'s are independent given the value of $\epsilon_t$. Hence,

\begin{equation}
Var(\sum_{i=1}^n Y_i|\epsilon_t) = \sum_{i=1}^n Var(Y_i|\epsilon_t) \nonumber 
\end{equation}
\begin{equation}
\label{eq:variance_light}
=nN^2 \frac{\sigma_{\gamma}^2}{\gamma^2}2 (p_0+p_0(1-p_0)\epsilon_t)^2 (1-(p_0+p_0(1-p_0)\epsilon_t))^2. \nonumber
\end{equation}
where the last equality is resulted by using $p_0+p_0(1-p_0)\epsilon_t$ as $p_0$ in~(\ref{eq:variance}) and neglecting $N$ relative to $N^2$.  By neglecting higher order terms of $\epsilon_t$, we obtain
\begin{equation}
\label{eq:variance_conditional}
E(Var(Y|\epsilon_t))=nN^2 p_0^2(1-p_0)^2 \frac{\sigma_{\gamma}^2}{\gamma^2}.
\end{equation}
On the other hand, we have
\begin{align}
\label{eq:expected_conditional}\nonumber
Var(E(\sum_{i=1}^n Y_i|\epsilon_t )&=Var\sum_{i=1}^n N(p_0+p_0(1-p_0)\epsilon_t)\\
&=N^2 n^2 p_0^2(1-p_0)^2 \sigma_t^2.
\end{align}
From~(\ref{eq:variance_conditional}) and~(\ref{eq:expected_conditional}) and using the conditional variance, we obtain
\begin{equation}
\label{eq:variance_final}
Var(Y)=nN^2(\frac{\sigma_{\gamma}^2}{\gamma^2}+n\sigma_t^2)^2 p_0^2(1-p_0)^2.
\end{equation}
Note that $\sigma_t^2$ (variance of $\epsilon_t$) is related to $\frac{\sigma_{\gamma}^2}{\gamma^2}$ through~(\ref{eq:variance_output}). With the same argument as in the transmitter case, we may approximate $Y$ with a Normal random variable. Hence, the output of the node $R$ would be in the form
\begin{equation}
\label{eq:receiver_output}
Y=nNp_0+\epsilon_Y,
\end{equation}
where $\epsilon_Y$ is a zero mean normal distribution with variance $Var(Y)$ given in~(\ref{eq:variance_final}). Note that, the first term in~(\ref{eq:receiver_output}) is the noiseless signal and the second one is a Normal additive noise.
In order to calculate the capacity per channel use from $T$ to $R$, we should obtain the optimized distribution of $p_0$ which maximizes $I(p_0;Y)$; the mutual information between the input and the output. This, in turn, gives the optimized distribution for $A_0$ through~(\ref{eq:steady_state}).

To proceed, we observe that in practice, $A_0$ cannot take any value. Hence, we assume the maximum achievable concentration is equal to $A_{max}$. This corresponds to probability $p_{max}=\frac{\gamma A_{max}}{\gamma A_{max}+\kappa}$ via~(\ref{eq:steady_state}). This maximum probability is due to the maximum power used by the transmitte. By using more power, the transmitter can increase the maximum concentration of molecules at the vicinity of the receiver node and increase $p_{max}$. Therefore, we obtain the optimized distribution for $p_0$ over the interval $ [0 \;\; p_{max}]$ and calculate the capacity based on $p_{max}$. 

The structure of the noise in~(\ref{eq:receiver_output}) is complicated since the noise power depends on the signal itself. Hence, we resort to use the numerical method of Blahut-Arimoto algorithm (BA) to obtain the optimal distribution for $p_0$ and its corresponding capacity.
\begin{figure}
\vspace{-1.1in}
\includegraphics[width=.48\textwidth]{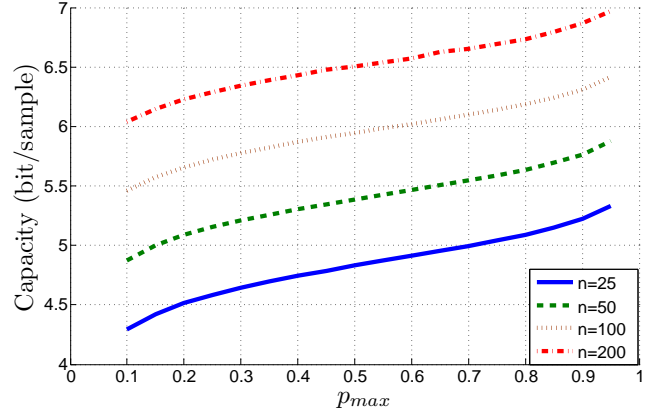}
\vspace{-1in}
\caption{Capacity (bits per sample) versus maximum trapping probability $p_{max}$ for different numbers of bacteria in a node. }
\vspace{-.17in}
\label{fig:capacity}
\end{figure}
 Equation~(\ref{eq:variance_final}) implies that the noise power is at its maximum at $p_0=\frac{1}{2}$ and goes to zero when $p_0$ approaches to either zero or one. Hence, we expect that the distribution of $p_0$ should take values closer to $0$ and $p_{max}$ with a higher probability. The results from the algorithm confirms this fact and the distribution has local maximums at $0$ and $p_{max}$.  
Results for the capacity (in bits per sample) with respect to $p_{max}$ for different numbers of bacteria in the nodes is shown in Fig.~\ref{fig:capacity}. In this setup, we assume $N=50$, $\frac{\sigma_{\gamma}^2}{\gamma^2}=.1$.
As we observe from the plot, the capacity increases when we increase $p_{max}$ or the number of bacteria $n$. Moreover, the convexity of the plots change at $p_{max}=\frac{1}{2}$. The reason for it is that after at this point, the variance of the noise starts to decrease. Note that the maximum achievable capacity is limited even if the transmitter used infinite power to make $p_{max}=1$. 
 In practice, $N$ and $n$ are very large. However due to the exponential growth of the simulation time, we only computed the capacity for small values of $N$ and $n$. 
\vspace{-0.1in}
\section{M-ary Modulation}
\label{sec:modulation}

The analysis in the previous section was based on the assumption that any continuous values of the concentration less than $A_{max}$ can be produced and received by the nodes.  In practice, we may use only a  finite discrete number of levels of molecular concentrations.
In this section, we consider M-ary modulation and study the the information exchange rate and the corresponding achieved error rate. The range of the input is determined by $p_{max}$. Two factors influence the signaling performance: the number of levels of concentration and the choices for the values of those levels. 
We consider the scenario in which $m$ symbols to be chosen with uniform spacing from the interval $[0\quad p_{max}]$. The $i^{th}$ symbol level would correspond to $p_{max} \frac{i}{m-1}, 0\leq i\leq m-1$. We show by $p_{e,i}$ the probability of error in the detection of $i^{th}$ symbol. Hence, the total probability of error is equal to $p_e=\sum_{i=0}^{m-1} w_i p_{e,i}$, where the weights $w_i$ associated with the $m$ symbols must be obtained .

\begin{figure}
\vspace{-1.2in}

\includegraphics[width=.48\textwidth]{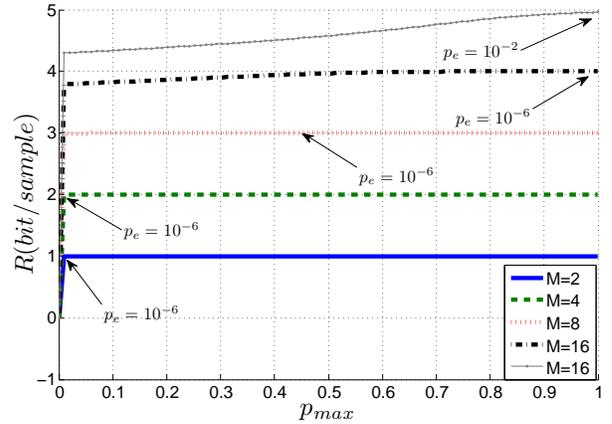}
\vspace{-1.2in}
\caption{The information rate versus the maximum power of the transmitter for different M-ary schemes}
\vspace{-.2in}
\label{fig:modulation}
\end{figure}

 We assume the error to occur when the detected symbol passes the half way from the previous or the next symbol. As observed in~(\ref{eq:variance_final}), the variance of the noise, and hence $p_{e,i}$ depends on the chosen symbol $i$. Therefore, we have
\begin{equation}
\label{eq:probability_error}
p_{e,i}=1-Pr(\frac{-p_{max}}{2(m-1)} \leq \epsilon_{Yi}  \leq \frac{p_{max}}{2(m-1)}), 
\end{equation}
where $\epsilon_{Yi}$ comes from a $\mathcal{N}(0,\sigma_i^2)$ where $\sigma_i^2$ can be computed by replacing $p_0$ with $\frac{i}{m-1}p_{max}$ in~(\ref{eq:variance_final}). As discussed in the previous section, variance of the noise is the smallest when the input is closets to $0$ or $1$. Hence, it is intuitive to choose larger weights for the inputs closer to these two points. In our scheme, we use the weights from the optimal distribution calculated by the Blahut-Arimoto algorithm. In Fig.~\ref{fig:modulation}, we have shown the rate of information for different M-ary modulations versus the power of the transmitter. In this setup, again we have chosen $N=50$ and $\sigma
_1^2=.1$. In addition, the number of bacteria in a node is chosen to be $n=100$.  As shown by the plot, reliable communication (i.e., $p_e=10^{-6}$) is feasible for $M=2, 4, 8, 16$ and the required power is shown as well. For larger number of symbols, reliable communication is not possible as for the case of $M=32$, the least error rate (by maximizing the $p_{max}$) would be $10^{-2}$. There, smaller error rates can be achieved by increasing either $n$ or $N$ (or both).

\vspace{-.05in}
\section{Conclusion}
\label{sec:conclusion}
In this paper, we studied the molecular communication between two nodes that contain populations of engineered bacteria. The error in the molecular production by the transmitter and the probabilistic nature of the reception of molecules at the receiver contribute to the noise in the communication. We studied the theoretical limits of the information transfer rate for different number of bacteria per node and different power levels. We observed that capacity increases with the number of bacteria in the nodes. Finally, we analyzed the rates and reliabilities in M-ary modulation. We observed that for a fixed number of bacteria per node and the number of ligand receptors, reliable communication is not possible for large $M$, even with increasing the input power.

\vspace{-.1in}

\bibliographystyle{IEEEtran}
\bibliography{ITW2012}
\end{document}